\documentclass[aps,nofootinbib]{revtex4}
\usepackage{epic}
\usepackage{eepic}
\newcommand{\be}{\begin{equation}}
\newcommand{\ee}{\end{equation}}
\newcommand{\bea}{\begin{eqnarray}}
\newcommand{\eea}{\end{eqnarray}}

\begin{document}

\baselineskip=.50cm

\title{On Winding Branes and Cosmological Evolution \\ of Extra
Dimensions in String Theory}
\author{Ali Kaya}
\email[e-mail:]{kaya@gursey.gov.tr}
\affiliation{Feza G\"{u}rsey Institute,\\
\c{C}engelk\"{o}y, 81220, \.Istanbul, Turkey}
\date{\today}
\begin{abstract}
\baselineskip=.40cm 
We consider evolution of compact extra dimensions in cosmology and try
to see whether wrapped  branes can prevent the expansion of the internal
space. Some difficulties of Brandenberger and Vafa mechanism for
decompactification are pointed out. In both pure Einstein and dilaton
gravities, we study cosmology of winding brane gases in a continuum
approximation. The energy momentum tensor is obtained by coupling the
brane action to the gravity action and we present several exact
solutions for various brane configurations. T-duality invariance of
the solutions are established in dilaton gravity. Our results indicate
that phenomenologically the most viable scenario can be realized when
there is only one brane wrapping over all extra dimensions.   
\end{abstract}
\maketitle
\section{Introduction}
String/M theory has a number of fascinating features as a unified
theory of all interactions. However, consistency of the theory 
requires existence of extra dimensions. Observationally, the sizes of
these dimensions are much smaller than the size of our perceived
universe. On the other hand, standard cosmology  tells us that the
universe started out very small (possibly close to Planck
length) and after various cosmological eras grew to its size
today. Therefore, one of the main problems of string/M theory applied
to cosmology is to determine why the extra dimensions remained
comparatively small during this cosmological evolution? 

In string theory, a remarkable mechanism to answer this
question was proposed in \cite{bv}. In that paper, Brandenberger and Vafa (BV)
first assumed that all dimensions were compact and started out very
small. Then, they noted that the winding states of the string would
like to prevent expansion since increasing the volume increases the
energy of these states. In thermal equilibrium, one expects a decrease
in the number of winding modes during expansion. Since winding number
is topologically conserved, only the strings having opposite
orientations can cause unwinding. However, such an interaction
takes place when the world-sheets of two such strings intersect and
this happens most efficiently in three spatial dimensions. 
Therefore in a three dimensional subspace winding strings can be
annihilated keeping them in thermal equilibrium with the rest of the
string modes and this results a decompactification. On the other
hand, strings winding other directions fail to annihilate each other,
fall out of equilibrium and stop the expansion. 

In \cite{abe} (see also related work on the subject of brane gas cosmology
\cite{eski1}-\cite{bgson}), this idea has been generalized to include higher
dimensional extended objects. Generically, two $p$-branes 
most efficiently intersect in at most $2p+1$ spatial
dimensions. Therefore, in nine dimensions (required by string
theory) all winding modes of $p$-branes with $p>3$ are
annihilated. Branes having smaller dimensions dictate a hierarchy of
sizes during expansion. First, 3-branes allow seven dimensions to become
large. Then, a five dimensional subspace is chosen by membranes. And
finally strings permit only a three dimensional subspace to grow. 

It is possible to point out three difficulties with the above
scenario. The first one is that the question of which dimensions are
picked up for expansion is left unanswered. Presumably,
quantum or thermal fluctuations determine this randomly.
However, as discussed in \cite{bv}, unwinding interactions
are very short ranged since this can only happen when the
world-volumes  physically intersect. Moreover in BV mechanism three
dimensions are picked up when the strings fail to intersect
each other effectively i.e. when the size of the
universe exceeds a critical value. Therefore, it is plausible that 
at the time of decompactification the universe contains many ``causal
patches'' with respect to unwinding interactions and different
dimensions become large in different patches.

The second difficulty is that, in principle, two $p$-branes with different
values of $p$ may find each other to change the winding number. 
Similar interactions are known to exist in string/M theory.
For instance, in the presence of background Ramond-Ramond fluxes,
Myers effect \cite{my} causes a point-like graviton to expand into a
spherical D3-brane. It is also well known that a D-brane can absorb
closed strings so that the quantum numbers (like the
winding number)  are encoded in the D-brane world-volume fields. Therefore, it
seems reasonable to expect that in nine spatial dimensions strings may
efficiently interact with seven or eight  dimensional branes to change the
winding number. This causes, without any difficulty, all winding
states to remain in thermal equilibrium with the rest of the modes.  

Finally, as discussed in \cite{biz}, a uniform distribution of 
winding $p$-branes along a transverse direction cause it to expand. 
Therefore, for more than one brane wrapping different
dimensions one has to take into account this effect. Consider, for
instance, a toy model of strings with two  compact dimensions, named $x$
and $y$. The strings winding $x$ look like point particles with
respect to $y$. Homogeneity requires a uniform distribution of such
strings in $y$ direction and this forces $y$ to expand (see Fig. 1).
In this case one has to compare this expansion to the contraction
forced by strings winding $y$ to decide whether decompactification occurs.

\begin{figure}[htb]
\begin{center}
\setlength{\unitlength}{0.00087489in}
\begingroup\makeatletter\ifx\SetFigFont\undefined%
\gdef\SetFigFont#1#2#3#4#5{%
  \reset@font\fontsize{#1}{#2pt}%
  \fontfamily{#3}\fontseries{#4}\fontshape{#5}%
  \selectfont}%
\fi\endgroup%
{\renewcommand{\dashlinestretch}{30}
\begin{picture}(2936,1734)(0,-10)
\put(-1063.805,777.646){\arc{3957.286}{5.7942}{6.6641}}
\put(-977.909,819.273){\arc{4414.582}{5.8693}{6.6430}}
\put(-397.000,874.500){\arc{3966.387}{5.8500}{6.7164}}
\put(244.250,874.500){\arc{4027.751}{5.8570}{6.7094}}
\put(165,875){\ellipse{314}{1664}}
\path(143,42)(2078,42)
\path(2348,42)(2348,402)
\path(2378.000,282.000)(2348.000,402.000)(2318.000,282.000)
\path(2348,42)(2708,42)
\path(2588.000,12.000)(2708.000,42.000)(2588.000,72.000)
\path(143,1707)(2078,1707)
\path(323,1167)(2258,1167)
\path(323,852)(2258,852)
\path(323,582)(2258,582)
\path(1133,1707)(1133,1257)
\blacken\path(1103.000,1377.000)(1133.000,1257.000)(1163.000,1377.000)(1103.000,1377.000)
\path(1133,42)(1133,492)
\blacken\path(1163.000,372.000)(1133.000,492.000)(1103.000,372.000)(1163.000,372.000)
\path(503,1257)(503,1617)
\blacken\path(533.000,1497.000)(503.000,1617.000)(473.000,1497.000)(533.000,1497.000)
\path(503,537)(503,177)
\blacken\path(473.000,297.000)(503.000,177.000)(533.000,297.000)(473.000,297.000)
\path(1898,537)(1898,132)
\blacken\path(1868.000,252.000)(1898.000,132.000)(1928.000,252.000)(1868.000,252.000)
\path(1898,1257)(1898,1572)
\blacken\path(1928.000,1452.000)(1898.000,1572.000)(1868.000,1452.000)(1928.000,1452.000)
\put(2393,492){\makebox(0,0)[lb]{\smash{{{\SetFigFont{12}{14.4}{\rmdefault}{\mddefault}{\updefault}$y$}}}}}
\put(2798,87){\makebox(0,0)[lb]{\smash{{{\SetFigFont{12}{14.4}{\rmdefault}{\mddefault}{\updefault}$x$}}}}}
\end{picture}}
\end{center}
\caption{The closed strings winding along $x$ and $y$ directions (the
torus is identified along $x$-direction). Strings winding $x$ 
force $y$ to expand. One has to compare this expansion to the
contraction forced by the strings winding $y$.}
\end{figure}
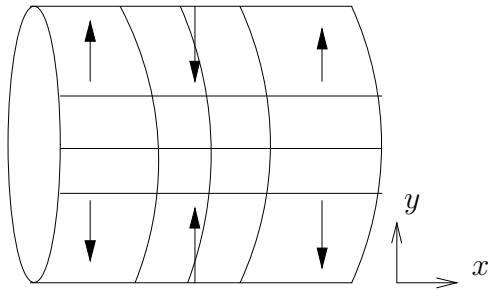

It is possible to avoid the first two complications mentioned above by
simply assuming that the observed three dimensional space is topologically
non-compact. This, one may hope, to be fixed by an unknown
stringy consistency condition. In this case, there is no problem
for these dimensions to expand. During this expansion, the density
of branes winding compact dimensions start to decrease in the three
dimensional subspace. At some point, these branes will be
diluted enough so that they generically will not be able to  find each
other to intersect. This will cause winding modes to fall out of
equilibrium and one hopes this may stop the expansion of the extra
dimensions.

In this paper, our aim is to analyze the third point raised
above more carefully. Assuming that the observed dimensions 
are topologically non-compact and the remaining ones form a
flat torus, we study toy cosmological models involving various winding
brane configurations.  

In \cite{biz}, two non-intersecting branes wrapping different internal
dimensions has been considered to compare the relative strengths of
the expansion and contraction forced by $p$-branes in pure Einstein gravity. 
Interestingly, it was found that whether the compact dimensions 
are getting large or small  depends on the dimension of the observed
space. Namely, for the two brane case, the internal dimensions are stabilized
when the observed space is three dimensional. Above three dimensions the
compact space contracts and below three it expands. As we will
see, there is a similar behavior for intersecting branes. 

The main  framework of \cite{biz} was the pure Einstein gravity. In
this paper, we extend the formalism to 
dilaton gravity. The dilaton played a crucial role in the development
of string cosmology. Especially, the large-small scale symmetry of
string theory called T-duality, which can only be established in the
presence of dilaton, has been argued to resolve the initial big bang
singularity in a natural way. The vacuum expectation value of the dilaton also
determines the string coupling constant and string theory enjoys a
strong-weak coupling symmetry called S-duality. We also discuss the
action of S and T-duality transformations in our cosmological context. 

For a single $p$-brane the energy momentum tensor can be
determined by coupling the brane action to the Einstein-Hilbert
action. This is a standard technique which has been used to determine
the geometry outside a macroscopic $p$-brane, see for instance
\cite{duff}. The energy momentum tensor obtained in this way  
supports a delta function singularity at the position of the brane along
the transverse directions. The delta function is smoothed out if one
assumes a gas of such $p$-branes in a continuum approximation \cite{biz}.  

Let us also emphasize that, in this paper we 
ignore the fluctuations of the brane coordinates and other 
world-volume fields.  This is a {\it classical} approximation to the
brane dynamics.  Therefore, our considerations can only be trusted at
some {\it late} time after the big bang when the temperatures are low
enough so that the brane fields are not exited.  

The organization of the manuscript is as follows. In section \ref{II},
we review the main formalism. In section \ref{III}, we study 
general aspects of a cosmology with non-intersecting winding branes in
pure Einstein gravity and argue that one should have at most two branes
to prevent cosmological expansion of the internal space. We also
construct solutions for intersecting $p$-branes. In section \ref{IV},
we  consider the dilaton gravity  and  present cosmological solutions
for single and intersecting D-branes . We also discuss T-duality
invariance of the solutions. In the last section, we summarize our
findings and speculate on a scenario based on our results.

\section{The Main Formalism}\label{II}

Let us consider the following $d$-dimensional metric which is of interest in
cosmological applications 
\be\label{met}
ds^2=-e^{2A}dt^2+\sum_{i}e^{2B_i}dx^idx^i.
\ee
Here the metric functions $A$ and $B_i$ depend only on time $t$. Using
$t$-reparametrization invariance one can set $A=1$. This is
the most preferred gauge choice in the literature. However,
there is another very useful gauge condition in which the Ricci tensor
simplifies considerably. It is easy to verify that, imposing $A=\sum_i
B_i$, the Ricci tensor takes the form \footnote{In our conventions, the
connection one-forms and curvature two-forms are given by
$de^{\hat{\mu}}+\omega^{\hat{\mu}}{}_{\hat{\rho}}\wedge
\omega^{\hat{\rho}}{}_{\hat{\nu}}=0$ and
$R^{\hat{\mu}}{}_{\hat{\nu}}=d\omega^{\hat{\mu}}{}_{\hat{\nu}}+
\omega^{\hat{\mu}}{}_{\hat{\rho}} \wedge\omega^{\hat{\rho}}{}_{\hat{\nu}}$.   
The Rici tensor is defined by $R_{\mu\nu}=R^{\rho}{}_{\mu\rho\nu}$.}
\bea\label{ricci}
R_{\hat{0}\hat{0}}&=&e^{-2A}(-A''+A'^2-\sum_iB_i'^2),\\
R_{\hat{i}\hat{j}}&=&e^{-2A}B_i''\,\delta_{ij}, \nonumber
\eea
where the hatted indices refer to the orthonormal frame $(e^A
dt,e^{B_i}dx^i)$ and $'$ denotes differentiation with respect to
$t$. We impose this gauge in solving the field equations
and then switch to the proper time coordinate after we obtain the
metric functions. 

Generically, one can consider the following energy momentum tensor for
matter
\be
T_{\hat{\mu}\hat{\nu}}=\textrm{diag}\,(\rho,p_{\hat{i}}),\label{energy1}
\ee
where $p_{\hat{i}}=\omega_{i}\,\rho$ with constant $\omega_i$. Energy
momentum conservation $\nabla_\mu T^{\mu\nu}=0$ gives
\be\label{omega}
\rho=\rho_0\,\,\exp\left[-\sum_{i}(1+\omega_i)B_i\right]\, ,
\ee
where $\rho_0$ is a constant. 

In string frame the dynamics of the metric and the dilaton are
governed via the action 
\be
S_{bulk}=\frac{1}{\kappa^2}\int d^{10}X\,\sqrt{-g_s}\,\, e^{-2\phi}\,\left[R_s+4(\nabla\phi)^2\right],\label{str}
\ee
where $\kappa^2=(2\pi)^7l_s$ and $l_s$ is the string length. 
We assume $4+6$ splitting of the space-time coordinates $X^\mu$
where the first three spatial coordinates are non-compact and the
other six  are topologically $S^1$. Ignoring the world-volume fields,
the action for a single $p$-dimensional D-brane is given 
\be
S_p=-2T_p\int d^{p+1}\xi\,e^{-\phi}\,\sqrt{-\gamma_s},\label{Dstr}
\ee
where $\xi^\alpha=(\tau,\sigma^a)$ are world-volume coordinates,
$\gamma^s_{\alpha\beta}$ is the pull back of the string metric 
and $T_p$ is the D-brane tension which can be
fixed by the string length. The complete effective action is 
\be
S=S_{bulk}+S_p\,.\label{tot}
\ee
We note that the string coupling constant $g_s$ is given by $e^\phi$.

The action (\ref{str}) is invariant under the S-duality
transformations 
\bea
\phi&\to&-\phi\label{S},\\
g^s_{\mu\nu}&\to& e^{-\phi}g^s_{\mu\nu}.\nonumber
\eea
The total action (\ref{tot}) is {\it not} invariant under
(\ref{S}). However, the transformed action can simply be interpreted
as a fundamental (S-dual) brane coupled to the bulk fields. While
the coupling of original D-brane to the bulk is inversely
proportional to the string constant $g_s$, the coupling of the
S-dual brane is proportional to $g_s^{(1-p)/2}$ \footnote{To be more
precise, type IIA string theory is mapped to M-theory
under S-duality. For instance type IIA D2-branes are related to 
membranes in eleven dimensions. On the other hand, type IIB theory is
self-dual; D1-branes are mapped to fundamental strings, D3-branes are
self-dual and D5-branes are mapped to magnetically charged 5-branes of
IIB theory. In this paper, we ignore all these stringy phenomena
(which are of course very important in developing string
cosmology) and see (\ref{S}) as a symmetry of the effective action
(\ref{str}).}. 

To see the action of T-duality we take a special configuration where
the world-volume coordinates are identified with some of the space-time 
directions. We also assume that the space-time metric is diagonal and
there is a Killing direction named $x$. Then the {\it total} action
(\ref{tot}) is invariant under the T-duality transformations 
\footnote{Actually, to establish the {\it exact} T-duality invariance
of string theory along a Killing coordinate, one should assume that
the isometry is compact. Here, we simply consider the invariance of
the  effective action and non-compactness of an isometry is not
a problem.} 
\bea
\phi&\to&\phi-\frac{1}{2}\ln (g^s_{xx}),\label{T}\\
g^s_{xx}&\to&1/g^s_{xx},\nonumber
\eea
where $x$ is identified as an additional brane coordinate if it is not
along the brane directions, or otherwise it is ignored as a brane
coordinate. Thus under a T-duality transformation a $p$-dimensional
D-brane can be mapped to a $(p-1)$ or $(p+1)$-dimensional D-brane. 
Note that T-duality of the effective action can be established
in the presence of D-branes and not the S-dual fundamental branes.   

In solving field equations, we find it convenient to work in the
canonical Einstein frame which is defined by 
\be\label{ein}
g^{s}_{\mu\nu}=e^{\phi/2}g_{\mu\nu}.
\ee
In terms of the Einstein metric $g_{\mu\nu}$, the total action
(\ref{tot}) can be written as 
\be
S=\frac{1}{\kappa^2}\int d^{10}X\,\sqrt{-g}\,\,\left[R-\frac{1}{2}(\nabla\phi)^2\right]-2T_p\int d^{p+1}\xi\,e^{a_p\phi}\,\sqrt{-\gamma},\label{eins}
\ee
where $a_p=(p-3)/4$ and $\gamma_{\alpha\beta}$ is the pull back of
the Einstein metric. 

From (\ref{eins}), one can determine the energy momentum tensor for a single
D-brane and the dilaton 
\bea\label{Tp}
\sqrt{-g}\,T_p^{\mu\nu}&=&-T_{p}\,\int
d^{p+1}\xi\,e^{a_p\phi}\,\sqrt{-\gamma}
\,\gamma^{\alpha\beta}\partial_\alpha X^\mu\partial_\beta
X^\nu\,\delta[X-X(\xi)],\\
\sqrt{-g}\,T^{\phi}_{\mu\nu}&=&\frac{1}{2}\nabla_\mu\phi\nabla_\nu\phi-\frac{1}{4}g_{\mu\nu}(\nabla\phi)^2\label{Tphi}
\eea
so that the field equations can be written as \footnote{One should
also consider the field equations for the map $X^{\mu}(\xi)$. These
are all satisfied identically for brane configurations where it is
possible to identify the world-volume coordinates with some of the space-time
coordinates.}  
\bea \label{eins1}
R_{\mu\nu}&-&\frac{1}{2}g_{\mu\nu}R=\kappa^2T^p_{\mu\nu}+T^\phi_{\mu\nu},\\
\nabla^2\phi&=&2a_p T_p \kappa^2\int
d^{p+1}\xi\,e^{a_p\phi}\,\frac{\sqrt{-\gamma}}{\sqrt{-g}}
\,\delta[X-X(\xi)].\label{dil}
\eea
To work in pure Einstein gravity one can set $a_p=0$ and ignore the
dilaton. 

As it is clear from the above formulas, for a single D-brane there is
a generic delta function singularity at the position of the brane
along the transverse directions. In a cosmological setting, it is
natural to take a gas of such D-branes in a continuum approximation 
(this is similar to assuming continuous charge distributions
in electromagnetism, although the unit of charge is quantized) and
this smooths out the singularity by an integration over transverse
dimensions \cite{biz}. We assume that the branes are uniformly
distributed and in this case one should replace the delta function by
a constant which is the average number of D-branes per unit
comoving transverse volume.

In the following sections, we take time dependent metric
functions and dilaton corresponding to certain brane
configurations and try to determine the cosmological evolution. Our
aim is to understand the main dynamical aspects and to find the
brane configurations which give contracting or stabilized
internal dimensions. 

\section{Cosmology of branes in Einstein gravity}\label{III}

In this section we concentrate on pure Einstein gravity and set
$a_p=\phi=0$. We first consider {\it non-intersecting} branes winding
different dimensions and argue (under some simplified assumptions) that
such a configuration cannot prevent cosmological expansion of all
internal dimensions unless there are only two branes. We then present
exact solutions for intersecting branes. In this section we also set
$\kappa^2=1$ and do not fix the dimension of the space-time.  

We take a $d$-dimensional metric of the form (\ref{met}) and consider
a uniform, continuum gas of toroidal $p$-branes, which can be
described by the map $t=\tau$, $x^a=\sigma^a$.  From (\ref{Tp}), the
energy momentum tensor can be written as (\ref{energy1}), where the
constants $\omega_i$ and $\rho_0$ are given by
\be
\omega_i=\begin{cases}{-1:\,\,\,\textrm{winding (brane) dimension},\cr\cr
\,\,\,\,0:\,\,\,\,\textrm{transverse dimension},}
\end{cases}\label{w}
\ee
$\rho_0=n\,T_p$ and $n$ is the number of branes per unit comoving
$(d-p-1)$-dimensional transverse volume. We assume that the first $m$ of 
$d-1$ spatial coordinates are non-compact playing the role of {\it
observed space}. Therefore $\omega_i=0$ for $i=1,..,m$. We also choose
$B_1=..=B_m\equiv B$ for isotropy.  

It is clear from (\ref{w})
that wrapped branes behave like pressureless dust with respect to the
directions they are distributed. On the other hand, they apply
negative pressure along the winding dimensions. One would naively
claim that negative pressure would cause these dimensions to expand as
in the case of vacuum energy domination during inflation. However, in
Einstein gravity, negative pressure does not necessarily imply
expansion. Indeed, even in de Sitter phase of the early universe, one
would have an exponential contraction with a negative Hubble
constant. Therefore, the situation highly depends on the initial
conditions.  

Adding $k$-different branes with dimensions $p_l$ ($l=1,..,k$), one has
$T_{\mu\nu}=\sum_l T^{(l)}_{\mu\nu}$. From the structure of the energy
momentum tensor, it is easy to see that
\be\label{tk}
g^{\mu\nu}T^{(l)}_{\mu\nu}=-(p_l+1)T^{(l)}_{\hat{0}\hat{0}}<0.
\ee
Therefore the trace of the energy momentum tensor is negative. 
Using (\ref{ricci}), the $(ij)$ components of the Einstein equations
(\ref{eins1}) for $i,j\leq m$ imply
\be\label{b}
(d-2)B''=-g^{\mu\nu}T_{\mu\nu}e^{-2A}.
\ee
Thus $B''>0$. On the other hand $(00)$ component of (\ref{eins1})
gives 
\be
A''-B''=\,T_{\hat{0}\hat{0}}\,\,e^{-2A}.
\ee
We have $A''>B''$ since $T_{00}>0$.

Now assume that the branes are not intersecting 
so that each internal dimension belongs to a single brane and we
have $d=1+m+\sum_l p_l$. We take the metric functions of the
same brane to be equal. Thus there are $k$ different functions 
multiplying extra dimensions which we label as $C^{(l)}$.  Using (\ref{b}),
Einstein equations (\ref{eins1}) for each brane direction give 
\be 
C^{(l)''} =-T^{(l)}_{\hat{0}\hat{0}}\,\, e^{-2A}+ B''
\ee
Multiplying the above formula with $(p_l+1)$, summing over $l$, 
using (\ref{tk}) and (\ref{b}) we obtain
\be\label{bad}
\sum_{l=1}^{k} (p_l+1) C^{(l)''} +(m-k-1)B''=0.
\ee

By integrating (\ref{bad}), one can fix one of the metric functions in
terms of the others up to the linear time dependent
terms. Generically, $(00)$  component of (\ref{eins1}) constraints the 
related undetermined integration constants. (The $(00)$ component is
special because of the gauge we choose where $g_{00}$ is determined by
the other metric functions.) As we will see, for many brane
configurations a power law ansatz for the metric is sufficient to
solve the Einstein equations. If this is the case then the linear
time dependent functions which arise in equations like (\ref{bad}) modify
this power law behavior. However, it is easy to see that in the $(00)$
component of (\ref{eins1}) the terms coming from these functions
have a different $t$ dependence. As a result they do not mix with the
other contributions, should sum up to zero separately and thus can be
ignored consistently. The only counterexample we find is in the context of
dilaton gravity where the power law ansatz fails and we will be forced
to keep these linear $t$ terms giving a different evolution (see the
metric (\ref{dsoln})). 

To proceed, let us now concentrate on power law solutions so
that the functions $A$, $B$ and $C^{(l)}$ are proportional to
$\ln(t)$ and ignore linear $t$ terms in (\ref{bad}). Then, for $m<k+1$, 
(\ref{bad}) implies that at least one of the functions $C^{(l)}$
should have the same sign with $B$, i.e. one of the
internal dimensions should behave like the observed ones and
expand \footnote{To see that the observed space expands, we start with
$A''>0$. This implies that original time coordinate $t$ and the proper
time are inversely proportional. Since $B''>0$, 
we have $B=-m\ln(t)$ for some positive $m$. Therefore
$e^B$ will be a decreasing function of $t$ and 
an increasing function of the proper time.}.
For $m=k+1$ either all the internal dimensions are stabilized or while 
some of them grow the others contract. For $m>k+1$, all extra 
dimensions can get small.

This has already been anticipated in \cite{biz}. Indeed, we
expect all internal dimensions to have the same behavior i.e. the
sign of $(m-k-1)$ determines whether the extra dimensions expand,
contract or stabilized. 

Since physically we have $m=3$, the above argument strongly suggests that
one should have {\it at most two} non-intersecting branes to prevent
cosmological expansion of the internal space. For instance, 
six internal dimensions cannot be held small by strings. Even for the
two brane case, which has been studied in detail in \cite{biz}, the
internal dimensions are stabilized. Including the states corresponding
to other fields and momentum modes of the branes, one would expect to
obtain enlarging extra dimensions. 

Having established the general picture for non-intersecting winding
branes, let us now consider some intersecting configurations to have a
more complete understanding. We first take two branes with dimensions $p+s$
and $q+s$ intersecting over an $s$-brane. The $d$-dimensional metric
can be written as 
\be \label{pqs}
ds^2=-e^{2A}dt^2+e^{2B}dx^idx^i+\sum_{l}e^{2C_l}dy^{a_l}dy^{a_l},
\ee
where $i=1,..,m$, $a_l,b_l=p,q,s$ for $l=1,2,3$ respectively and 
we again impose the gauge $A=mB+pC_1+qC_2+sC_3$. The first brane with
tension $T_1$ has the world-volume coordinates $(y^{a_1},y^{a_3})$
and the second brane with tension $T_2$ wraps over the torus
parametrized by $(y^{a_2},y^{a_3})$. Using (\ref{w}), the total 
energy momentum tensor corresponding to this configuration 
can be written as
\bea
T_{\hat{0}\hat{0}}&=&n_1T_1\exp[-mB-qC_2]+n_2T_2\exp[-mB-pC_1],\nonumber\\
T_{\hat{i}\hat{j}}&=&0,\nonumber\\
T_{\hat{a}_1\hat{b}_1}&=&-n_1T_1\exp[-mB-qC_2]\delta_{a_1b_1},\label{tpqs}\\
T_{\hat{a}_2\hat{b}_2}&=&-n_2T_2\exp[-mB-pC_1]\delta_{a_2b_2},\nonumber\\
T_{\hat{a}_3\hat{b}_3}&=&-n_1T_1\exp[-mB-qC_2]\delta_{a_3b_3}-n_2T_2\exp[-mB-pC_1]\delta_{a_3b_3},\nonumber
\eea
where $n_1$ and $n_2$ are the number of branes per unit respective
comoving transverse volumes. From (\ref{tpqs}), the
Einstein equations can be written as
\bea
A''-A'^2&+&mB'^2+pC_1'^2+qC_2'^2+sC_3'^2=C_3'',\nonumber\\
(d-2)B''&=&(p+s+1)\,F_1+(q+s+1)\,F_2,\nonumber\\
(d-2)C_1''&=&-(m+q-2)\,F_1+(q+s+1)\,F_2,\label{eins3}\\
(d-2)C_2''&=&(p+s+1)\,F_1-(m+p-2)\,F_2,\nonumber\\
(d-2)C_3''&=&-(m+q-2)\,F_1-(m+p-2)\,F_2,\nonumber
\eea
where
\bea
F_1&=&n_1T_1\exp[mB+2pC_1+qC_2+2sC_3], \\
F_2&=&n_2T_2\exp[mB+pC_1+2qC_2+2sC_3].
\eea 
Although it looks quite complicated, it is remarkable that
(\ref{eins3}) admits an exact power law solution. To determine the metric, 
let us first note that the last four equations in (\ref{eins3}) give 
\bea
(m-s-3)B+(p+s+1)C_1+(q+s+1)C_2&=&0,\label{bc}\\
B-C_1-C_2+C_3&=&0,\nonumber
\eea
where we ignore linear integration terms (see the discussion below
(\ref{bad})). Using (\ref{bc}), one can express the second and the
third equations of (\ref{eins3}) in terms of $B$ and $C_1$ alone. To
solve them, we assume  
\bea
B&=&b_1\ln(t)+\ln b_2, \label{ans}\\
C_1&=&c_1\ln(t)+\ln c_2,\nonumber
\eea
where $b_1$, $b_2$, $c_1$ and $c_2$ are constants. A straightforward
calculation shows that these constants can be determined
uniquely. Using (\ref{bc}) and the gauge choice for $A$, one can then 
determine all the metric functions. The first equation in
(\ref{eins3}) is then to be checked for consistency and it is 
satisfied identically. On the other hand, the constants
$b_2$ and $c_2$ should be determined to be non-zero which imposes
\be
pq(9-s)+4sp+4qs-mq(s-1)-mp(s+q-1)\not= 0.
\ee
This condition is obeyed for the physically important case of $m=3$.
All these calculations involve trivial algebraic manipulations which
can be performed by using a computer program. 

After solving all the unknown metric functions, one can switch to the proper
time coordinate (which we again denote by $t$) to get
\be \label{met3}
ds^2=-dt^2\,+\,R_m^2dx^idx^i\,+\,\sum_{l}R_l^2\,dy^{a_l}dy^{a_l},
\ee
where
\bea
\ln R_m &=& 2\Delta^{-1}[p+q+2pq+ps+qs]\,\ln(\alpha_m t),\nonumber\\
\ln R_1 &=& -2q\Delta^{-1}[m-s-3]\,\ln(\alpha_1 t),\label{fin1}\\
\ln R_2 &=& -2p\Delta^{-1}[m-s-3]\,\ln(\alpha_2 t),\nonumber\\
\ln R_3 &=& -\Delta^{-1}[2(m-2)(p+q)+4pq]\,\ln(\alpha_3 t).\nonumber
\eea
Here $\alpha_m$, $\alpha_l$ are dimensions-full constants 
determined by $T_1$, $T_2$, $n_1$, $n_2$ and the positive number 
$\Delta$ is given by
\be 
\Delta=pq(3+s)+m(1+s)(p+q)+mpq.
\ee
The values of $\alpha_m$ and $\alpha_l$ set the units of length for
each direction and they can be eliminated by scalings of
$x$ and $y$ coordinates. Note that this also modifies the energy
momentum tensor.  

Setting $s=0$, ignoring $R_3$ and the corresponding coordinates,
(\ref{met3}) becomes the metric constructed in \cite{biz} for two
non-intersecting branes. This is a consistency check on
both solutions. However, let us emphasize that setting spatial brane
dimensions to zero is {\it not} always possible. For instance one
cannot let $p=0$ further to obtain the solution for a single winding 
brane. To see this, we note that the trace of the energy momentum
tensor is proportional to $p+1$ and setting $p=0$ does not remove the
trace part.   

From (\ref{fin1}), one finds that the observed space expands and the
intersecting directions contract. On the other hand, the sign of
$(m-s-3)$  determines the fate of the relative transverse directions
\footnote{A coordinate is defined to be a relative transverse one 
if it belongs to a brane but transverse to another one.}. When $m<s+3$, these
dimensions expand. For $m=s+3$ they are stabilized and for $m>s+3$
they contract. Therefore, for the physically important case of $m=3$
one finds that  the intersecting dimensions diminish and the relative
transverse coordinates enlarge. Comparing with the non-intersecting
solution which gives stable extra dimensions, this
behavior suggests that the internal space can be seen as an  
elastic balloon; when some directions are forced to contract the
others tend to expand. 

Using (\ref{met3}), it is also possible to obtain the special
intersection where an $s$-brane is located inside a $(p+s)$-brane 
by setting $q=0$ in (\ref{fin1}) and ignoring the corresponding 
coordinates (as we pointed out this is not a trivial operation and one
should check the steps carefully). In this way, one can obtain the
following metric
\be \label{eski}
ds^2=-dt^2\,+\,(\alpha_m
t)^{\frac4m}dx^idx^i\,+\,dy^{a_1}dy^{a_1}\,+\,(\alpha_3 t)^{-\frac{4(m-2)}{m(s+1)}}dy^{a_3}dy^{a_3},
\ee
where $(y^{a_1},y^{a_3})$ and $(y^{a_3})$ are
the world-volume coordinates of $(p+s)$ and $s$-dimensional
branes, respectively, and we scale $y^{a_1}$ to set $\alpha_1=1$.
In the $p$-dimensional subspace spanned by $y^{a_1}$, the contraction
forced by the winding $(p+s)$-branes is compensated by the expansion
forced by the gas of $s$-branes. Ignoring this part, 
(\ref{eski}) is precisely the metric corresponding to 
$s$-dimensional winding branes which has been found in \cite{biz}.   

Finally, let us consider adding a third $(p+q+s)$-dimensional
brane to (\ref{met3}) winding over all extra dimensions. This is a
triple intersection. One should modify (\ref{tpqs}) and (\ref{eins3})
by adding terms coming from the third brane source. The rest of the
calculation can be carried out as follows. The last four 
equations in (\ref{eins3})  give a relation between $B$ and
$C_l$. Using this relation, one can eliminate one of the functions and
obtain three differential equations for three functions which can be
solved using an ansatz like (\ref{ans}). It is then possible
to determine all metric functions and the first equation in
(\ref{eins3}) is satisfied identically. In the proper time coordinate
the final result can be written as  
\be \label{eski2}
ds^2=-dt^2\,+\,(\alpha_m
t)^{\frac4m}dx^idx^i\,+\,dy^{a_1}dy^{a_1}\,+\,dy^{a_2}dy^{a_2}\,+\,
(\alpha_3 t)^{-\frac{4(m-2)}{m(s+1)}}dy^{a_3}dy^{a_3},
\ee
where we scale $y^{a_1}$ and $y^{a_2}$ to set
$\alpha_1=\alpha_2=1$. Ignoring these coordinates, this is again
precisely the metric for a gas of winding $s$-branes. Let us remind
that in (\ref{eski2}), there are three branes which has the
world-volume coordinates $(y^{a_1},y^{a_3})$, $(y^{a_2},y^{a_3})$ and
$(y^{a_1},y^{a_2},y^{a_3})$, respectively. It is remarkable
that, as in the case of two non-intersecting branes, in certain directions
the expansion and contraction forced by the branes cancel each other.
This suggests that it may be possible to establish a {\it
superposition} rule for wrapped branes and this might have important
cosmological applications. 

Before closing this section, we would like to address an important
issue: since the energy of the winding branes increases as the
internal dimensions expand, one may think that in the above solutions
the total energy is not conserved when this happens. This claim is not
correct since all energy momentum tensors
considered so far obey $\nabla_\mu T^{\mu\nu}=0$ which guarantees
conservation of energy. One may then wonder where this extra energy
comes from?  To answer this question, let us remind that we employ
an approximation where the branes are assumed to be continuously
distributed. Therefore, in addition to the winding energy one can talk
about energy corresponding to this continuous distribution. When the winding
energy increases the  energy of the distribution decreases and the
total energy remains constant.

How we should understand the flow between winding and distribution
energies in the real situation where the continuum approximation is
not valid. Since we assumed that the branes fall out of thermal
equilibrium and do not interact with each other, it seems odd to say
that the increase in the energy of a single winding brane is
compensated by a decrease in the energy of the others. To understand
this let us consider a two dimensional  cylinder with winding strings
on it (see Fig. 2). When the strings are dense enough they are able 
to prevent the radial expansion. However, as the cylinder undergoes a
linear cosmological expansion, the number of strings per unit length
decreases. In this case, winding strings may not be able to prevent
the radial expansion of the space in between them. Of course, whether
this happens is determined by the dynamical field equations. If 
it turns out to be the case, then the space whose radial expansion is
prevented by the strings on the cylinder becomes of measure zero in
time and one sees a radially expanding cylinder.

\begin{figure}[htb]
\begin{center}
\setlength{\unitlength}{0.00087489in}
\begingroup\makeatletter\ifx\SetFigFont\undefined%
\gdef\SetFigFont#1#2#3#4#5{%
  \reset@font\fontsize{#1}{#2pt}%
  \fontfamily{#3}\fontseries{#4}\fontshape{#5}%
  \selectfont}%
\fi\endgroup%
{\renewcommand{\dashlinestretch}{30}
\begin{picture}(5827,1389)(0,-10)
\put(-348.000,687.000){\arc{2340.000}{5.8884}{6.6780}}
\put(-438.000,687.000){\arc{2257.189}{5.8731}{6.6933}}
\put(-213.000,687.000){\arc{2340.000}{5.8884}{6.6780}}
\put(-78.000,687.000){\arc{2340.000}{5.8884}{6.6780}}
\put(113.250,687.000){\arc{2319.247}{5.8847}{6.6817}}
\put(2172.000,687.000){\arc{2257.189}{5.8731}{6.6933}}
\put(2847.000,687.000){\arc{2340.000}{5.8884}{6.6780}}
\put(3657.000,687.000){\arc{2340.000}{5.8884}{6.6780}}
\put(4197.000,732.000){\arc{2340.000}{5.8884}{6.6780}}
\put(3881.075,646.623){\arc{3875.613}{5.9051}{6.5924}}
\put(439,687){\ellipse{134}{900}}
\put(2779,665){\ellipse{226}{1304}}
\path(462,1137)(1182,1137)
\path(462,237)(1182,237)
\path(417,687)(12,687)
\blacken\path(132.000,717.000)(12.000,687.000)(132.000,657.000)(132.000,717.000)
\path(1317,732)(1587,732)
\blacken\path(1467.000,702.000)(1587.000,732.000)(1467.000,762.000)(1467.000,702.000)
\path(2757,1317)(2759,1316)(2762,1313)
	(2768,1308)(2778,1301)(2790,1292)
	(2806,1280)(2824,1267)(2845,1253)
	(2867,1238)(2890,1224)(2915,1209)
	(2940,1195)(2967,1182)(2995,1170)
	(3025,1160)(3057,1151)(3091,1143)
	(3126,1139)(3162,1137)(3201,1139)
	(3235,1145)(3263,1153)(3285,1164)
	(3300,1175)(3310,1188)(3315,1200)
	(3318,1214)(3320,1227)(3321,1240)
	(3324,1254)(3329,1266)(3339,1279)
	(3354,1290)(3376,1301)(3404,1309)
	(3438,1315)(3477,1317)(3513,1315)
	(3547,1311)(3577,1304)(3603,1295)
	(3623,1285)(3640,1274)(3652,1263)
	(3662,1251)(3669,1239)(3676,1227)
	(3682,1215)(3690,1203)(3700,1191)
	(3712,1180)(3729,1169)(3751,1159)
	(3777,1150)(3808,1143)(3844,1139)
	(3882,1137)(3919,1139)(3954,1143)
	(3983,1150)(4007,1159)(4025,1169)
	(4038,1180)(4047,1191)(4053,1203)
	(4056,1215)(4058,1227)(4061,1239)
	(4064,1251)(4070,1263)(4079,1274)
	(4093,1285)(4111,1295)(4136,1304)
	(4167,1311)(4202,1315)(4242,1317)
	(4279,1316)(4314,1312)(4346,1306)
	(4374,1298)(4398,1289)(4418,1280)
	(4434,1270)(4447,1259)(4457,1249)
	(4466,1238)(4475,1227)(4483,1216)
	(4491,1205)(4502,1195)(4514,1184)
	(4529,1174)(4548,1165)(4571,1156)
	(4597,1148)(4627,1142)(4659,1138)
	(4692,1137)(4730,1139)(4764,1145)
	(4790,1153)(4810,1163)(4824,1174)
	(4832,1186)(4837,1198)(4838,1211)
	(4838,1223)(4838,1236)(4839,1249)
	(4843,1261)(4850,1273)(4862,1285)
	(4879,1296)(4902,1305)(4930,1313)
	(4962,1317)(4994,1318)(5024,1315)
	(5049,1309)(5069,1302)(5083,1293)
	(5094,1283)(5100,1272)(5105,1261)
	(5108,1249)(5112,1238)(5117,1227)
	(5124,1216)(5136,1206)(5152,1197)
	(5175,1190)(5203,1184)(5238,1181)
	(5277,1182)(5313,1186)(5348,1192)
	(5382,1200)(5414,1210)(5444,1221)
	(5472,1233)(5499,1246)(5524,1259)
	(5549,1273)(5572,1287)(5594,1301)
	(5615,1315)(5633,1327)(5649,1338)
	(5661,1347)(5671,1354)(5677,1358)
	(5680,1361)(5682,1362)
\path(2802,12)(2803,14)(2806,17)
	(2812,23)(2820,32)(2831,44)
	(2845,58)(2861,74)(2879,92)
	(2899,110)(2921,129)(2944,147)
	(2968,164)(2995,180)(3023,195)
	(3054,209)(3088,220)(3125,229)
	(3165,235)(3207,237)(3246,235)
	(3282,230)(3315,223)(3344,213)
	(3369,203)(3388,191)(3404,178)
	(3417,165)(3427,152)(3436,138)
	(3443,124)(3451,111)(3459,97)
	(3469,84)(3481,71)(3495,58)
	(3514,46)(3536,36)(3562,26)
	(3592,19)(3624,14)(3657,12)
	(3695,15)(3728,22)(3754,32)
	(3772,45)(3783,60)(3789,75)
	(3790,91)(3788,108)(3784,125)
	(3781,141)(3780,158)(3781,174)
	(3787,189)(3800,204)(3820,217)
	(3848,227)(3884,234)(3927,237)
	(3961,236)(3995,231)(4028,225)
	(4058,217)(4084,207)(4108,197)
	(4128,186)(4145,174)(4160,162)
	(4174,149)(4186,137)(4197,124)
	(4208,112)(4220,100)(4234,87)
	(4249,75)(4266,63)(4286,52)
	(4310,42)(4336,32)(4366,24)
	(4399,18)(4433,13)(4467,12)
	(4510,15)(4546,22)(4575,32)
	(4595,44)(4609,59)(4616,74)
	(4619,89)(4619,105)(4617,121)
	(4615,137)(4614,153)(4616,168)
	(4622,184)(4633,198)(4650,212)
	(4674,223)(4703,232)(4737,237)
	(4769,237)(4799,234)(4825,228)
	(4849,220)(4868,209)(4883,198)
	(4896,186)(4905,173)(4913,160)
	(4921,147)(4928,134)(4936,121)
	(4946,108)(4958,96)(4973,85)
	(4991,74)(5013,66)(5039,60)
	(5068,57)(5097,57)(5132,63)
	(5161,73)(5182,87)(5196,103)
	(5204,120)(5206,137)(5205,155)
	(5202,173)(5199,191)(5198,209)
	(5201,226)(5210,243)(5226,258)
	(5250,270)(5282,279)(5322,282)
	(5353,280)(5385,275)(5417,268)
	(5447,258)(5477,247)(5505,234)
	(5533,220)(5561,205)(5587,189)
	(5613,172)(5638,155)(5663,139)
	(5685,123)(5706,107)(5724,94)
	(5740,82)(5752,73)(5761,66)
	(5767,61)(5770,58)(5772,57)
\end{picture}}
\end{center}
\caption{Strings winding a two dimensional cylinder. As the cylinder
expands horizontally the number of strings per unit length
decreases. As a result, the strings may not be able to prevent the radial
expansion of the space in between them.}
\end{figure}
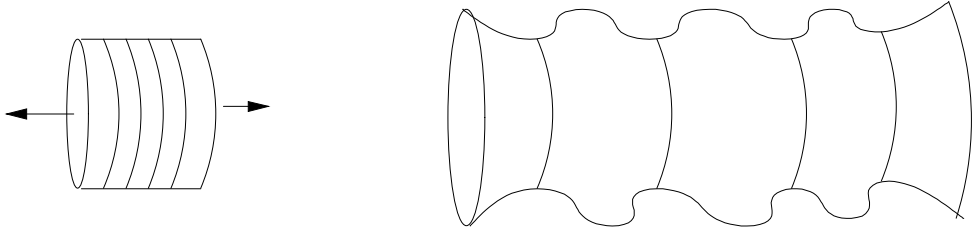

\section{Adding Dilaton}\label{IV}

In this section, we construct solutions for winding branes in
dilaton gravity. Adding dilaton, one discovers new physical
phenomena to study. For instance, it is possible to define string and
Einstein frame metrics and generically one would expect to observe
different cosmological behavior in each frame. As we will see,
T-duality invariance plays a key role in a cosmology D-brane gases in
dilaton gravity. 

Let us first consider a single winding D-brane, which can be described
by  following ten dimensional metric  
\be \label{dilp}
ds^2=-e^{2A}dt^2+e^{2B}dx^idx^i+e^{2C}dy^{a}dy^{a},
\ee
where $i,j=1,..,m$ and $a,b=1,..,p$, $m+p=9$ and $A=mB+pC$. 
As before, we take $p$-dimensional D-branes wrapping the
internal circular dimensions $y^a$ and uniformly distributed in the
observed space spanned by $x^i$. We carry out the computation in
Einstein frame and then switch to string frame.  
Using (\ref{Tp}) and (\ref{Tphi}), it is easy to
write down the {\it total} energy momentum tensor (we set $\kappa^2=1$)
\bea 
T_{\hat{0}\hat{0}}&=&n\,T_p\,e^{[-mB-a_p\phi]}+\frac{1}{4}\phi'^2\,e^{-2A},
\nonumber\\
T_{\hat{i}\hat{j}}&=&\frac{1}{4}\phi'^2\,e^{-2A},\label{tmunu}\\
T_{\hat{a}\hat{b}}&=&-n\,T_1\,e^{[-mB-a_p\phi]}\delta_{ab}+\frac{1}{4}\phi'^2\,e^{-2A}\delta_{ab},\nonumber
\eea
where $a_p=(p-3)/4$. Due to the non-trivial coupling in (\ref{eins})
only the total $T_{\mu\nu}$ is conserved provided that the equations of
motion are satisfied. The Einstein and the dilaton field equations can be
written as
\bea
A''&-&A'^2+mB'^2+pC'^2+\frac{1}{2}\phi'^2=C'',\nonumber\\
B''&=&\frac{p+1}{m+p-1}n\,T_p\exp[mB+2pC+a_p\phi],\label{eins4}\\
C''&=&-\frac{m-2}{m+p-1}n\,T_p\exp[mB+2pC+a_p\phi],\nonumber\\
\phi''&=&-2nT_p\,a_p\exp[mB+2pC+a_p\phi].\nonumber
\eea
The last equation indicates that the dilaton can effectively be seen
as an additional internal dimension. This is not surprising since the
circular coordinate, on which one can relate M-theory to type IIA
string theory by a compactification, is parametrized by the vacuum
expectation value of the dilaton. 

We first attack these equations without fixing the values of $m$, $p$
and $a_p$ to see the effect of adding dilaton in Einstein
gravity. Later we focus on the case implied by string theory. In
accordance with our main strategy, using the last three 
equations in (\ref{eins4}) one can determine $C$ and $\phi$ in terms
of $B$. The second equation in (\ref{eins4}) can then be used to solve
$B$ by an ansatz like (\ref{ans}) and this determines all metric
functions. Switching to the proper time coordinate one can finally obtain
\bea \label{met5}
ds^2_E&=&-dt^2\,+\,R_m^2dx^idx^i\,+\,R_p^2\,dy^{a}dy^{a},\\
e^\phi&=&R_\phi\nonumber
\eea
where
\bea
\ln R_m &=& \frac{2(p+1)}{m(p+1)+2(m+p-1)a_p^2}\,\ln(\alpha t),\nonumber\\
\ln R_p &=& -\frac{2(m-2)}{m(p+1)+2(m+p-1)a_p^2}\,\ln(\alpha t),\label{fin2}\\
\ln R_\phi &=& -\frac{4a_p(m+p-1)}{m(p+1)+2(m+p-1)a_p^2}\,\ln(\alpha t).\nonumber
\eea
For $a_p=0$, the dilaton decouples and we find the single winding brane
solution of \cite{biz}. Adding dilaton modifies the power law but does
not change the physically expected behavior where the observed space
expands and the compact space contracts due to winding of branes.

In (\ref{fin2}), the constant $\alpha$ should be determined to be
non-zero. One can check that this implies
\be\label{neg}
a_p^2\not=\frac{m-mp+4p}{2(m+p-1)}.
\ee
When (\ref{neg}) is not satisfied the power law ansatz which has been
used to solve the above differential equations cannot be
used. 

Returning now to string theory, we have $m+p=9$ and $a_p=(p-3)/4$. It
is remarkable that for these values (\ref{neg}) is {\it not}
obeyed. Therefore, (\ref{met5}) cannot be used to describe the
cosmology of winding branes in string theory. To obtain the proper
solution, we first followed our main strategy. Namely we
first determined $C$ and $\phi$ in terms of $B$ neglecting linear
$t$ terms. After solving $B$ from the second equation in (\ref{eins4})
we obtained all metric functions. However, a quick analysis showed
that the first equation in (\ref{eins4}) cannot be satisfied in this
way.  

Therefore, this time one has to keep linear $t$ terms in obtaining relations
between $B$, $C$ and $\phi$. Keeping them, (\ref{eins4}) gives
\bea
C&=&\frac{p-7}{p+1}B\,+\,c\,t,\label{1}\\
\phi&=&\frac{4(3-p)}{p+1}B\,+\,d\,t,\nonumber
\eea
where $c$ and $d$ are constants. Using (\ref{1}), the second equation in
(\ref{eins4}) becomes 
\be
8\,B''=n\,T_p\,(p+1)\,\exp[(2\,p\,c+a_p\,d)t],\label{2}
\ee
which can be solved as
\be
B=\frac{(p+1)nT_p}{8(2\,p\,c+a_p\,d)^2}\exp[(2\,p\,c+a_p\,d)t]+bt,\label{3}
\ee
where $b$ is a
constant. Using (\ref{1}), (\ref{3}) and the gauge condition for $A$,
one finds that (\ref{eins4}) are satisfied provided that
\be\label{cons}
-\frac{16\,b\,(2\,p\,c+a_p\,d)}{p+1}-p^2\,c^2+p\,c^2+\frac{d^2}{2}=0.
\ee
It is possible to eliminate $b$, $c$ or $d$ by  
scaling $t$ appropriately. Since two remaining undetermined constants
are constrained by (\ref{cons}), there is a single parameter solution
space. However, we would like to construct a solution which is invariant
under T-duality transformations. To determine the constants for  
this solution, we first switch to string frame. From
(\ref{ein}), the string frame metric functions are defined by
\bea
A_s&=&A+\frac{\phi}{4}=-\frac{12}{p+1}B\,+\,(pc+\frac{d}{4})t, \nonumber\\
B_s&=&B+\frac{\phi}{4}=\frac{4}{p+1}B+\frac{d}{4}\,t,\label{sframe}\\
C_s&=&C+\frac{\phi}{4}=\frac{4}{p+1}B+(c+\frac{d}{4})t.\nonumber
\eea
For invariance under a T-duality transformation (\ref{T}), one should
have $B_s=-C_s$, and this imposes $c=-d/2$. On the other hand, by scaling
$t$, one can set $c=2/(p+1)$ and (\ref{cons}) can be used to
find $b$. This gives
\bea
A_s&=&3f+t,\nonumber\\
B_s&=&f-t/3,\\
C_s&=&-f+t/3,\nonumber\\
\phi&=&-(p-3)f+\frac{p-6}{3}t,\nonumber
\eea
where $f=\lambda\,\,e^{3t}$
and $\lambda=nT_p/18$. Defining a new time coordinate by $t\to e^t$,
we obtain the {\it string frame} metric and the dilaton
\bea
ds_S^2&=&-e^{6\lambda t^3}dt^2\,+\,t^{-\frac{2}{3}}e^{2\lambda
t^3}dx^idx^i\,+\,t^{\frac{2}{3}}e^{-2\lambda t^3}dy^ady^a,\label{dsoln}\\
e^\phi&=&t^{\frac{(p-6)}{3}}e^{-(p-3)\lambda t^3}.\nonumber
\eea
One can easily see that applying a T-duality transformation along $x$
direction gives a solution for a gas of $(p+1)$-dimensional
D-branes. Similarly, a T-duality along $y$ direction reduces the brane
dimension by one. It is also straightforward to calculate the S-dual
solution from (\ref{S}).

One would claim that (\ref{dsoln}) gives exponential scaling functions
for the observed and the internal spaces. However, note that $t$ is
not the proper time coordinate. It is indeed possible to show that
when expressed in terms of the proper time $\tau$, which can be defined as
\be\label{prop}
d\tau=e^{3\lambda t^3}dt,
\ee 
(\ref{dsoln}) is close to a power low solution. To see this, one can
integrate (\ref{prop}) so that as $t\to-\infty$,
$\tau\to0$.  When $|t|$ is large (\ref{prop}) gives
$\tau\sim e^{3\lambda t^3}/(9\lambda t^2)$. On the other hand when $t$
is small, one finds $\tau\sim t$. The Hubble parameters with respect
to $\tau$ can be calculated as
\be
H_m=-H_p=\frac{d}{d\tau}\ln R_m=\frac{(9\lambda
t^3-1)}{3t}\,\,e^{-3\lambda t^3}.
\ee
When $t$ is very large and very small one finds $H_m\sim \pm 1/(3\tau)$,
respectively,  which indicates a cosmological evolution obeying a
power law $R_m\sim \tau^{\pm1/3}$.  

At $\tau=0$ ($t=-\infty$), the observed space has zero size and the
scale function of the compact space diverges. As $\tau$ increases, the
observed space expands and the internal space contracts where the
metric functions have power law dependences on the proper time.
However, at some $\tau$ (given by $t=0$) the scale factor of the
observed space diverges and the compact space has zero size. After
that time the observed space starts to diminish and the compact space
enlarges. This period is followed by an expansion of the observed
space and the contraction of the internal space, which is 
again close to a power law. 

In string theory, (\ref{dsoln}) can be trusted when the
curvatures and the string coupling $g_s$ are small. For example, one
cannot trust the geometry near $t=0$ since the curvatures are large. 
Similarly, for $p>3$, one cannot use (\ref{dsoln}) near $t\sim-\infty$
since string coupling diverges. This shows that (\ref{dsoln}) can be
trusted when $|t|$ is large. Moreover, for $p>3$, $t$ should take
positive values and for $p<3$ it should be negative. In this range of
validity, we discover the physically expected behavior; that is the
observed space expands and the extra dimensions contract. 

Having studied the single D-brane case, let us now consider
an intersecting brane configuration. Assume that in the six
dimensional compact space there are $(p+s)$ and $(q+s)$-dimensional 
D-branes with tensions $T_1$ and $T_2$ intersecting over an
$s$-dimensional submanifold, where $p+q+s=6$. As before, we carry out
calculations in Einstein frame and then switch to the string frame.
The metric can be written as
\be \label{dpqs}
ds^2=-e^{2A}dt^2+e^{2B}dx^idx^i+\sum_{l}e^{2C_l}dy^{a_l}dy^{a_l},
\ee
where $i=1,2,3$, $a_l,b_l=p,q,s$ for $l=1,2,3$ respectively and 
we again impose the gauge $A=mB+pC_1+qC_2+sC_3$. 
The first brane wraps on the torus parametrized by
$(y^{a_1},y^{a_3})$ and the second brane has the toroidal world-volume
coordinates $(y^{a_2},y^{a_3})$. Using (\ref{Tp}) and (\ref{Tphi}) the
total energy momentum tensor can be calculated as 
\bea
T_{\hat{0}\hat{0}}&=&n_1T_1\exp[-mB-qC_2+a_p\phi]+n_2T_2\exp[-mB-pC_1+a_q\phi],\nonumber\\
T_{\hat{i}\hat{j}}&=&0,\nonumber\\
T_{\hat{a}_1\hat{b}_1}&=&-n_1T_1\exp[-mB-qC_2+a_p\phi]
\delta_{a_1b_1},\label{diltpqs}\\
T_{\hat{a}_2\hat{b}_2}&=&-n_2T_2\exp[-mB-pC_1+a_q\phi]
\delta_{a_2b_2},\nonumber\\
T_{\hat{a}_3\hat{b}_3}&=&-n_1T_1\exp[-mB-qC_2+a_p\phi]\delta_{a_3b_3}
-n_2T_2\exp[-mB-pC_1+a_q\phi]\delta_{a_3b_3},\nonumber
\eea
where $a_p=(p+s-3)/4$ and $a_q=(q+s-3)/4$. From (\ref{diltpqs}), 
the field equations (\ref{eins1}) and (\ref{dil}) can be written as
\bea
A''&-&A'^2+mB'^2+pC_1'^2+qC_2'^2+sC_3'^2+\frac{1}{2}\phi'^2=C_3'',\nonumber\\
8\,B''&=&(p+s+1)\,F_1+(q+s+1)\,F_2,\nonumber\\
8\,C_1''&=&-(q+1)\,F_1+(q+s+1)\,F_2,\label{eins5}\\
8\,C_2''&=&(p+s+1)\,F_1-(p+1)\,F_2,\nonumber\\
8\,C_3''&=&-(m+q-2)\,F_1-(m+p-2)\,F_2,\nonumber\\
\phi''&=&-2a_pF_1-2a_qF_2\nonumber
\eea
where
\bea
F_1&=&n_1T_1\exp[mB+2pC_1+qC_2+2sC_3+a_p\phi], \\
F_2&=&n_2T_2\exp[mB+pC_1+2qC_2+2sC_3+a_q\phi].
\eea 
To solve these equations we first note that, up to the linear $t$ terms
which we ignore, (\ref{eins5}) implies
\bea
(1+q+s)C_2&=&sB-(p+s+1)C_1,\nonumber\\
(1+q+s)C_3&=&-qB+(q-p)C_1,\label{ex}\\
(1+q+s)\,\phi\,&=&-2sB+2(p-q)C_1.\nonumber
\eea
These relations can be used to express the second and the third
equations of (\ref{eins5}) in terms of $B$ and $C_1$ alone. To solve
them, one can use the ansatz (\ref{ans}) and see that 
all constants are uniquely fixed. From  (\ref{ex}) and gauge condition
$A=mB+pC_1+qC_2+sC_3$, all metric functions and the dilaton can be
determined. It is also possible to show that the first equation in
(\ref{eins5}) is satisfied identically. Switching to the proper time
coordinate we finally obtain the following solution in {\it Einstein frame}
\bea 
ds^2_E&=&-dt^2\,+\,(\alpha t)^{\frac{16+2s}{3(10-p-q)}}dx^idx^i\,+\,\
(\alpha t)^\frac{2s}{3(10-p-q)}(dy^{a_1}dy^{a_1}+ dy^{a_2}dy^{a_2})\,+\,
(\alpha t)^{-\frac{16-2s}{3(10-p-q)}}dy^{a_3}dy^{a_3},\nonumber\\
e^{\phi}&=&(\alpha t)^{\frac{-4s}{3(10-p-q)}},\label{met6}
\eea
where $\alpha$ is a {\it positive} dimension-full number. 

The metric in (\ref{met6}) shows that the observed space expands
(close to a power law corresponding to pressureless dust) and the
intersecting dimensions contract. On the other hand, the relative
transverse directions also enlarge. Comparing with the intersecting
brane solution in pure Einstein gravity for $m=3$, we see
that adding dilaton modifies the power law but does not change the
main physical behavior. 

Let us now switch to the {\it string frame} metric defined by
(\ref{eins}). In terms of the proper time coordinate we find
\bea \label{met7}
ds^2_S&=&-dt^2\,+\,(\alpha t)^{\frac{8}{(12-p-q)}}\,dx^idx^i\,+\,
(dy^{a_1}dy^{a_1}+ dy^{a_2}dy^{a_2})\,+\,
(\alpha t)^{-\frac{8}{(12-p-q)}}\,dy^{a_3}dy^{a_3},\\
e^{\phi}&=&(\alpha t)^{\frac{-2s}{(12-p-q)}}.\nonumber
\eea
As in the previous cases, we see that the observed space expands and the
intersecting directions contract. On the other hand, the
dilaton stabilizes expansion of the relative transverse space
parametrized by $(y^{a_1},y^{a_2})$. This is remarkable compared to
the behavior we observed in (\ref{met6}). 

One can easily see from (\ref{met7}) that a T-duality transformation 
along $x$ shifts $s\to s+1$ and decreases the dimension
of the observed space by one. This gives a solution describing
$(p+s+1)$ and $(q+s+1)$-dimensional branes intersecting over an
$(s+1)$-dimensional internal space. On the other hand, a T-duality
along $y^{a_1}$ gives a metric for the intersection
of $(p+s-1)$ and $(q+s+1)$-dimensional branes over an $s$-dimensional
space. Transformations along $y^{a_2}$ and $y^{a_3}$ can be
interpreted similarly. Note that the value of $p+q$ is not altered 
under the T-duality. 

It is possible to argue that T-duality invariance of string theory
can be used to analyze more general brane configurations. For the
moment, let us assume that there are many number of different D-branes
winding the compact internal space. One can group the space-time
coordinates into three disjoint sets: the observed, the relative
transverse and the intersecting ones.  If a coordinate is common to all
branes it is an intersecting dimension. If it belongs to a brane but
it is transverse to another one, it is a relative transverse
coordinate. If there are no  branes wrapping on it, it is one of the
observed coordinates. One can easily see that the set of relative
transverse coordinates is invariant under T-duality
transformations. On the other hand, the observed coordinates are
interchanged with the intersecting ones. Since physically one expects
that the observed space expands, T-duality invariance dictates that
the intersecting dimensions should contract. On the other hand, the
relative transverse coordinates should be stabilized since one can
interchange two of them by a T-duality. Therefore, 
T-duality invariance of the effective action implies
that the internal dimensions {\it cannot} expand. This is contrary to the
behavior we observed in Einstein gravity and shows the importance of
T-duality in string cosmology.

\section{Conclusions}

In this paper, we study cosmology of winding branes in both pure Einstein
and dilaton gravities. The energy momentum tensor is obtained by
coupling the brane action to the gravity action. We take a uniform gas
of such branes and utilize a continuum approximation. The brane
fluctuations and the world-volume fields are ignored in our analysis.
Generically, we see that the internal dimensions wrapped by $p$-branes
tend to contract. On the other hand, the gas of such winding
$p$-branes force the transverse dimensions to expand. 

In pure Einstein gravity, cosmological evolution in the presence of
single and two non-intersecting branes has been determined in
\cite{biz}. The single brane wrapping over all extra dimensions cause
them to diminish. On the other hand for two branes
winding different internal dimensions, the expansion and the
contraction is stabilized when the observed space is three
dimensional. We argue that if there are three or more non-intersecting
winding branes the cosmological expansion of the
internal space cannot be prevented. We also consider intersecting
brane configurations. In this case, we find that the common brane 
directions contract while the relative transverse directions tend to
enlarge or stay stabilized.  

Adding dilaton to pure Einstein gravity, we see that the above
conclusions do not change appreciably. For instance, two intersecting
branes give expanding relative transverse directions. 

On the other hand, the dilaton has a remarkable effect in
string frame; it stabilizes the expansion of the
relative transverse coordinates. We argued that, T-duality invariance
of the effective action can be used to 
determine the main cosmological behavior for more general
systems. Namely, one expects to find an expanding observed space,
stable relative transverse directions and contracting intersecting
dimensions.  

Is it possible to determine which brane configurations can play a role
in keeping the extra dimensions small and which ones should be
eliminated?  To answer this question let us point out that
till now we have ignored two important issues. The first one is that
we have only considered winding modes of the branes. To have a more
complete scenario one should include effects of other sources, like
the momentum modes, which are expected to force all dimensions to
expand. Secondly, there is now considerable evidence
indicating that the universe underwent an exponentially growing period,
called inflation, forced by a positive cosmological constant.   
Inflationary paradigm is successful in explaining away the basic
shortcomings of standard  cosmology and its predictions are confirmed
with the astrophysical observations so far. Phenomenologically one
expects 60 or 70 e-foldings during inflation. As discussed in
\cite{biz}, with a positive cosmological constant the brane terms have
negligible effects and can be ignored. Thus, 
internal space was also subject to an exponential growth. 
Assuming that the compact dimensions started out at about
Planck length, one finds that their size grew to about $10^{-5}m$
after inflation. This indicates that the configuration
dominating the subsequent cosmological evolution should force {\it
all} internal dimensions to contract and the results of this
paper show that this can only be realized when there is a single brane
wrapping over all extra dimensions.

It is then possible to speculate on the following scenario. The ten
dimensional universe was born in a hot, dense state at about the
Planck length. For some reason (which hopefully will be found by string/M
theory) three spatial coordinates were started to be non-compact
and the remaining ones parametrize a six dimensional compact internal
space. The universe then underwent an inflationary period where all
dimensions were subject to exponential growth. The inflation was
followed by an expansion close to a power law. During this period
winding branes could not prevent the cosmological expansion but they
slowed it down. However, in a short time all winding D$p$-branes with
$p<6$ were annihilated. One can claim that they could all be absorbed
by D6-branes since they were wrapping over all extra
dimensions leaving no room for other branes. 
Then the internal space started to contract and this
might explain the hierarchy between the sizes of the extra and the
observed dimensions.  

To see that such a scenario can produce phenomenologically viable
numbers for the size of the internal space, let us ignore all other
sources and consider the cosmological evolution after the
inflation when there is a single brane wrapping over all extra
dimensions. The scale functions of the observed space $R_o$ and the
internal space $R_i$ are related by $R_i=R_o^{-1/(p+1)}$ in pure
Einstein gravity \cite{biz} (where $p$ is the number of extra
dimensions) and $R_i=1/R_o$ in dilaton gravity.
The estimates obtained in \cite{biz} indicate that in
pure Einstein gravity $p>1$ is not viable phenomenologically \footnote{The
analysis of \cite{biz} ignored one possible caveat that in reality the
extra dimensions do not grow in the same amount as the observed
space during inflation due to winding branes. However, a numerical
study shows that, although the situation may depend on the
values of the brane tension and the energy density driving inflation,
the difference cannot be accommodated in this way \cite{tong}.}. To be
more precise, let us assume that the diameter of presently
observable universe (our current Hubble volume), 
which is of the order of $10^{27}m$, was
about $10^{-5}m$ at the end of inflation. Therefore, the scale
function of the observed space for this time interval can be estimated
to be $R_o\sim 10^{32}$. Using this number, the scale function of the
internal space becomes $R_i\sim 10^{-16}$ in pure Einstein gravity
for $p=1$ and $R_i\sim 10^{-32}$ in string cosmology. This gives the
current size of the internal space to be $10^{-21}m$ in Einstein gravity and
$10^{-37}m$ in string theory. Doing the same exercise for $p=6$ in
pure Einstein gravity one finds unacceptable estimates. It is
remarkable that the dilaton modifies the relation between the scale
functions in such a way that even for six dimensional branes wrapping
over all extra dimensions one can obtain phenomenologically viable
estimates. 

We conclude by nothing that in string theory it is possible to
obtain cosmological backgrounds with contracting extra dimensions
without introducing winding brane sources explicitely (see for
instance S-brane solutions \cite{s1}-\cite{sson}). In these solutions
there are non-zero form-fluxes and this is a different way to describe
branes in a supergravity context. It would be interesting to find a
connection between the solutions presented in this paper and the other
cosmological backgrounds related to S-branes.   

\begin{acknowledgments}
We would like to thank N.S. Deger and T. Rador for useful
discussions. 
\end{acknowledgments}

\end{document}